# Corrosion Risk Estimation for Heritage Preservation: An Internet of Things and Machine Learning Approach Using Temperature and Humidity

Reginald Juan M. Mercado[1], Muhammad Kabeer[2,3], Haider Al-Obaidy[4], Rosdiadee Nordin[2,5]

[1] GTek Enterprise, Project 2, Quezon City 1102, Philippines

[2] Faculty of Engineering and Technology, Sunway University, No. 5, Jalan University, Bandar Sunway 47500, Malaysia

[3] Department of Computer Science, Federal University Dutsinma, Katsina, Nigeria

[4] Department of Information and Communications Engineering, College of Information Engineering, Al-Nahrain University, Baghdad, Iraq

[5] Future Cities Research Institute, Sunway University, No. 5, Jalan University, Bandar Sunway 47500, Malaysia

## Abstract

Proactive preservation of steel structures at culturally significant heritage sites like the San Sebastian Basilica in the Philippines requires accurate corrosion forecasting. This study designed and developed an Internet of Things (IoT) hardware connected with LoRa wireless communications that monitors heritage building with steel structures. From the three-year dataset generated from the IoT system, we further develop a machine learning framework for predicting atmospheric corrosion rates using minimal environmental data, specifically temperature and relative humidity measurements. The analysis employs three ensemble regression models Random Forest, Gradient Boosting, and XGBoost trained on temporally structured data with rigorous time-series validation to ensure realistic forecasting conditions. All models demonstrated exceptional predictive performance, with Random Forest emerging as the optimal choice ($R^2$ = 0.9913, MAE = 0.89 μm/year, RMSE = 7.86 μm/year). Gradient Boosting and XGBoost also achieved strong results with $R^2$ values of 0.9902 and 0.9881, respectively. Notably, XGBoost demonstrated superior computational efficiency with a training time of only 2.62 seconds, significantly faster than Random Forest (102.92 seconds) and Gradient Boosting (78.45 seconds), while maintaining competitive accuracy. The sub-1 μm/year mean absolute error achieved by Random Forest represents exceptional precision for practical conservation applications. Deployed via a Streamlit dashboard with ngrok tunneling for public access, this framework provides real-time corrosion monitoring and actionable preservation recommendations. This minimal-data approach offers a scalable, cost-effective solution for heritage sites with limited monitoring resources, demonstrating that advanced regression techniques can extract highly accurate corrosion predictions from basic meteorological data, enabling proactive preservation strategies for culturally significant structures worldwide without requiring extensive sensor networks.

## 1.0 Introduction

Cultural heritage structures are invaluable assets that represent historical, social, and cultural identity. Their preservation is increasingly jeopardized by environmental stressors, such as climate change, pollution, and natural aging processes. Structural health monitoring (SHM) has thus become an essential instrument for evaluating deterioration and informing preventive conservation measures. Heritage buildings, in contrast to modern infrastructure, pose distinct challenges owing to their intricate geometries, diverse material characteristics, and stringent conservation mandates. The reasons render the implementation of successful SHM methodologies both technically challenging and essential [1], while digital innovations such as context-aware recommender systems have also been explored to enhance heritage management [2].

Atmospheric corrosion of metals is a continual and inadequately monitored concern among the degradation mechanisms that jeopardize heritage structures. Landmarks like the San Sebastian Basilica, built using innovative steel technology, illustrate the susceptibility of heritage structures to corrosion, significantly affected by relative humidity, temperature variations, and exposure to pollutants. The visual evidence of rust on the Basilica's structural components depicted in Figure 1 highlights the detrimental effects of extended environmental exposure and inadequate oversight. Conventional methods for assessing corrosion risk, such as dose response functions (DRFs), have been extensively utilized to measure the cumulative impact of climatic factors and air pollution on material degradation. Originally established through extensive European initiatives and subsequently expanded to areas like China, DRFs are crucial for producing quantitative forecasts and guiding conservation programs [3]. Nonetheless, their reliance on pollutant concentration data, frequently inaccessible or inconsistently monitored at heritage sites, constitutes a significant restriction.

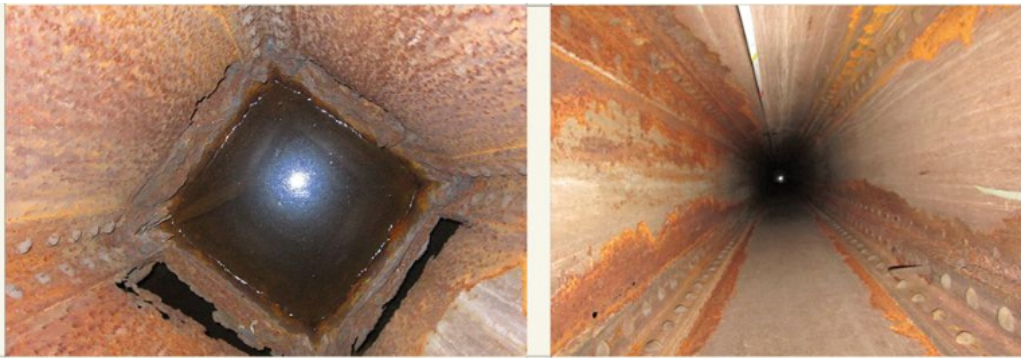

*Figure 1 Inside Column Showing Water and Large Holes Caused by Rusting*

Recent advancements highlight both the efficacy and limitations of pollutant-dependent frameworks. Ríos-Rojas *et al.* [4] created a Bogota-specific Durability Risk Factor for carbon steel that integrated relative humidity, temperature, $SO_2$, and particulate matter, demonstrating the influence of urban microclimates on corrosion patterns. Paterlini *et al.* [5] corroborated ISO 9223 DRFs in Milan and Bonassola, demonstrating a strong correlation between projected and observed corrosion rates for carbon and galvanized steels. Research conducted in subtropical regions, including the Canary Islands, indicates a consistent underestimation of corrosion severity, as ISO 9223 models do not adequately account for microclimatic variability and necessitate the introduction of new categories, such as CX, for extreme exposure situations [6]. These findings demonstrate the twofold challenge: DRFs offer mechanistic foundations and international comparability, yet their reliance on pollutants and geographical variations limit their scalability.

Concurrent progress in conservation-aligned monitoring and data science offers opportunity to overcome these obstacles. Franzoni and Bassi [7] established the viability of incorporating non-destructive, sensor-equipped ceramic plugs for the surveillance of growing damp in historic masonry. The CULTCOAST project in Norway highlighted the significance of site-specific monitoring, while recognizing the logistical and resource limitations associated with multi-sensor installations [8]. The integration of IoT, cloud, and edge computing at the systems level facilitates real-time environmental information derived from disparate data streams [9]. Outside of historic contexts, interpretable machine learning has demonstrated efficacy, [10] utilized LightGBM with SHAP analysis on reinforced concrete subjected to marine exposure, revealing the evolution of environmental feature significance during its service life. These works together indicate that combining physics-based corrosion knowledge with data-efficient machine learning frameworks can produce predictive algorithms that are less dependent on pollutant datasets while effectively capturing nonlinear environmental interactions.

Despite the strengths of DRFs and multi-sensor IoT frameworks, there is still a lack of scalable corrosion monitoring solutions that rely only on universally available climate data. This study responds directly to this need by advancing a minimal-data, computationally efficient corrosion forecasting framework. The contributions are:

- Designed and developed a custom LoRa-based IoT system for heritage monitoring, enabling long-range, low-power measurement of environmental parameters including temperature and humidity without compromising structural integrity.

- Presents a physics-based corrosion rate model calibrated against literature data to quantify the effects of temperature and humidity, demonstrating that accurate monitoring can be achieved using only minimal inputs enabling low-cost, scalable deployment.

- Predictive framework combining advanced regression models for continuous corrosion rate estimation with a binary alarm index that translates predictions into actionable risk classifications, enabling precise long-term forecasting and rapid conservation decision-making.

This study enhances existing DRF methodologies [3] by tackling their reliance on pollutants, thereby providing a scalable solution for heritage conservation and contemporary infrastructure management within resource limitations.

## 2.0 Related Works

### 2.1 Heritage Conservation and Environmental Monitoring

The conservation of cultural heritage structures necessitates comprehensive environmental monitoring, as these sites are especially susceptible to atmospheric stresses. DRFs, developed via European field exposure trials, have been beneficial in forecasting corrosion rates based on pollutant concentrations, temperature, and relative humidity [3]. Nonetheless, their dependence on pollutant monitoring constrains applicability in heritage contexts where multi-pollutant datasets are limited. Initiatives such as CULTCOAST in Norway underscore the necessity for site-specific monitoring while simultaneously exposing resource limitations that hinder extensive sensor implementation [8]. Recent reviews also emphasize the role of IoT-based SHM in extending the service life of civil and heritage structures through real-time monitoring and AI-enabled analysis [11].

Progress in non-invasive monitoring corresponds with conservation concepts advocating minimal intervention. Franzoni and Bassi [7] presented a sensorized ceramic plug for historic masonry, whilst [1] highlighted IoT-enabled monitoring as a scalable solution. Ficili *et al.* [9] demonstrated how the combination of IoT, cloud, edge, and AI can convert disparate sensor information into predictive intelligence applicable to both heritage and infrastructure. Complementing these approaches, innovative stress sensors embedded in mortar joints have been demonstrated to provide reliable monitoring of masonry structures without compromising material [12].

## 2.2 Atmospheric Corrosion Modeling: Machine Learning and Hybrid Approaches

The forecasting of atmospheric corrosion has progressively utilized data-driven methodologies. [13] devised a two-stage hybrid machine learning approach that integrates tree ensembles with neural networks to address small-sample corrosion datasets. [14] utilized random forest models using galvanic sensor data, identifying temperature, humidity, and rainfall as primary short-term determinants, while contaminants were shown to be secondary factors. [15] merged feature selection with support vector regression to improve dependability in long-term datasets.

Additionally, [10] illustrated interpretable ensemble learning for reinforced concrete subjected to maritime exposure, whereby the evolving feature importance elucidated the changing influence of temperature, humidity, and chloride ingress. These studies collectively demonstrate that machine learning can exceed the accuracy of decision rule frameworks while also improving interpretability, thereby enabling the prioritization of universally quantifiable factors for efficient monitoring.

## 2.3 Atmospheric Corrosion Models

Classical DRFs constitute the foundation of international corrosion evaluation, however their application differs by geography and environment. [4] formulated a Bogotá-specific Damage Risk Function for carbon steel that integrated relative humidity, temperature, sulfur dioxide, and particulate matter, demonstrating that microclimatic variability within urban areas influences localized corrosion risks. [5] corroborated ISO 9223 predictions through mass loss and electrochemical assessments in Milan and Bonassola, demonstrating consistency in urban environments (C2 class) but markedly elevated marine aggressivity (C4–C5). [6] revealed that ISO 9223 underestimates corrosivity in subtropical archipelagic regions such as the Canary Islands, where observed zinc and copper losses surpassed C5 levels, requiring the CX category for extreme conditions.

These investigations reveal a dual perspective, DRFs offer mechanistic and standardized corrosion predictions but, their reliance on pollutants and susceptibility to spatial heterogeneity impede scalability, especially in heritage contexts where pollutant monitoring is infrequent.

Notwithstanding methodological advancements, three fundamental gaps persist. Hybrid and ensemble models typically necessitate comprehensive contaminant and deposition data, which are unfeasible in heritage situations. Secondly, most of the research emphasizes industrial or coastal infrastructure, with insufficient adaptation to conservation contexts where minimal intervention is crucial. Third, although short-term monitoring elucidates corrosion dynamics, the translation of these insights into long-term projections appropriate for conservation decision-making remains unaddressed.

Further obstacles encompass challenges related to scalability, interoperability, and interpretability within IoT-AI systems [9]. Notably, few models investigate minimal-data frameworks that depend exclusively on universally accessible climate factors, despite evidence indicating that temperature and humidity are predominant in early-stage corrosion processes. Addressing this disparity necessitates the integration of physics-based models with interpretable machine learning, ensuring adherence to both scientific integrity and heritage conservation tenets.

### 3.0 Methodology

To advance the preservation of heritage structures, this project deploys an integrated pipeline that combines a LoRa-based IoT system with machine learning algorithms to enable continuous environmental monitoring and predictive corrosion analytics. *Figure 2* presents the end-to-end architecture, encompassing sensor deployment, data preprocessing, model training, and real-time deployment. This framework transforms raw environmental inputs into actionable insights, supporting evidence-based conservation strategies and adaptive risk management.

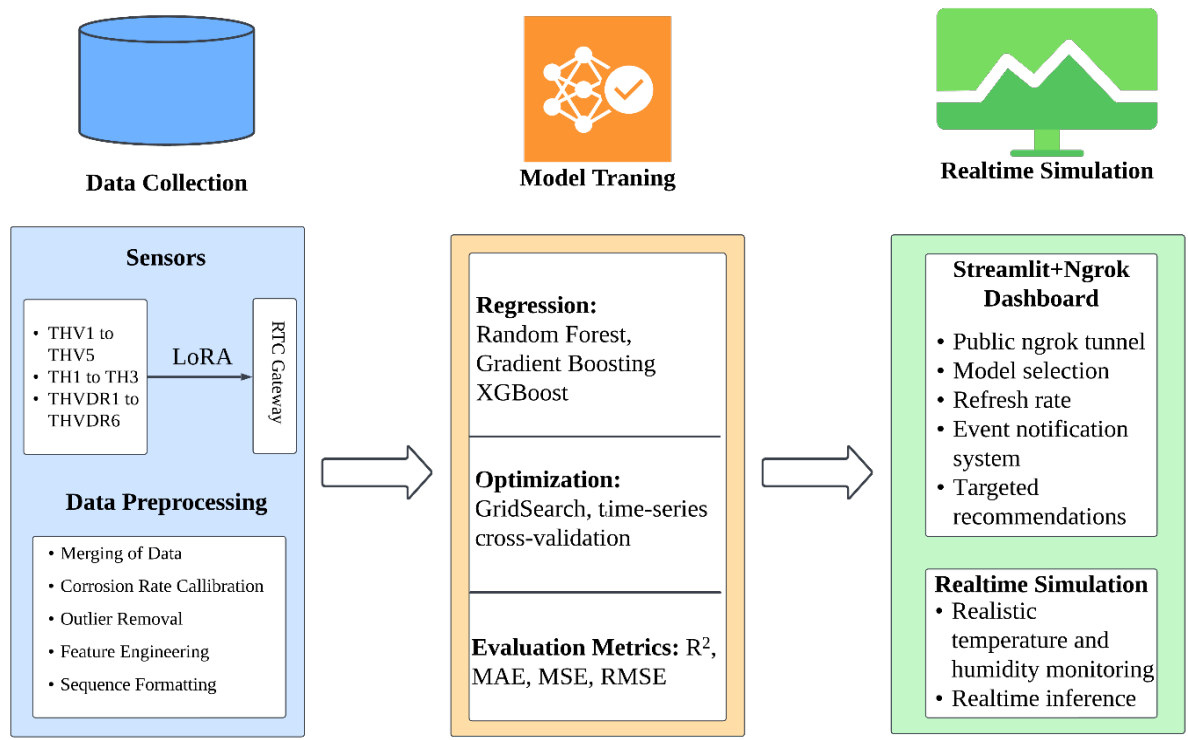

*Figure 2 Heritage Preservation Methodology*

### 3.1 Deployment of LoRa-based IoT System

Environmental monitoring at the San Sebastian Basilica was carried out using a wireless sensor network of 14 stations and one LoRa gateway as shown *Figure 3*. Three types of nodes were deployed: (i) TH nodes (temperature and humidity), (ii) THV nodes (temperature, humidity, wind speed), and (iii) THVDR nodes (temperature, humidity, wind speed, wind direction, rainfall). Indoor sensors were placed in critical locations *Figure 4*, while outdoor weather stations were positioned at doorways, the roof, and the tower to capture external climatic drivers *Figure 5*.

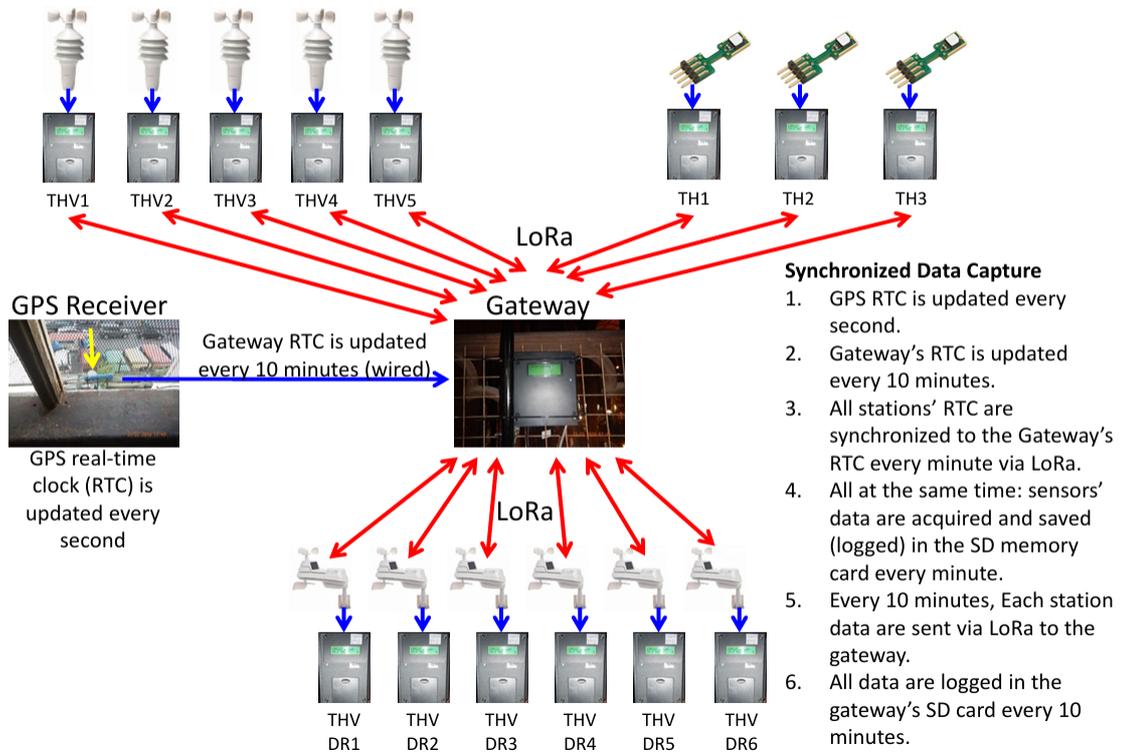

Figure 3 LoRa Environmental Monitoring System Operation

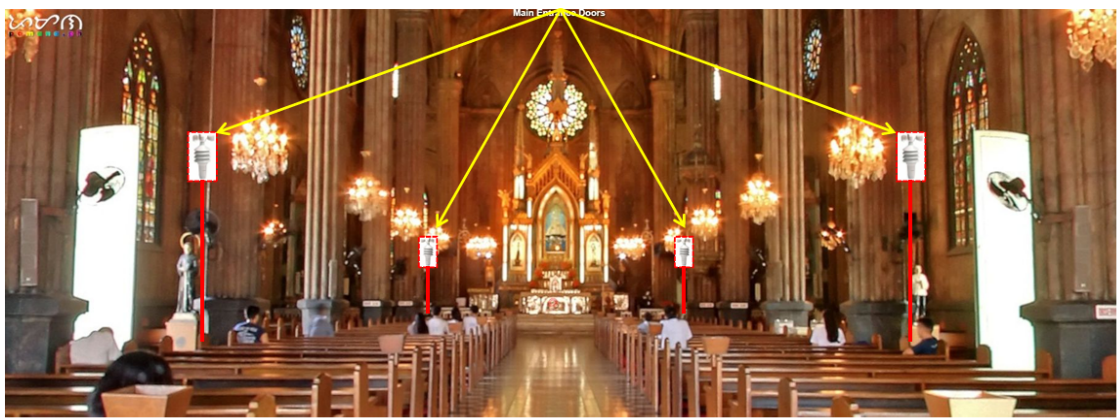

Figure 4 Indoor Ambience Sensors at Level 1 of the church

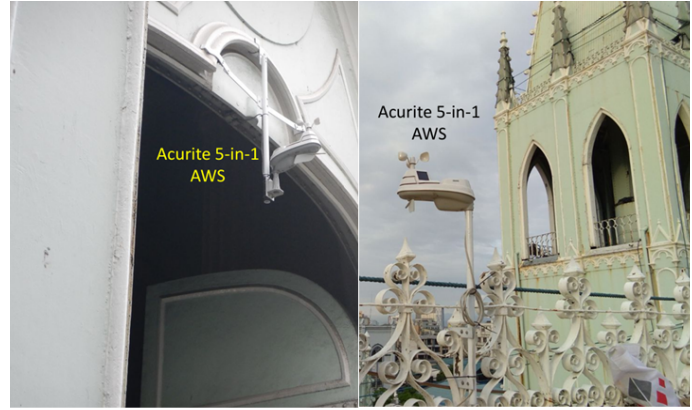

*Figure 5 Deployment of Sensor in Outdoor Locations*

The system architecture ensures synchronized sampling. A GPS receiver updates the real-time clock (RTC) every second, which in turn updates the gateway RTC every 10 minutes. Each station then synchronizes its clock with the gateway once per minute via LoRa. Data acquisition occurs simultaneously across all stations, with sensor values logged locally to SD cards each minute and transmitted to the gateway every 10 minutes. The gateway consolidates and stores all incoming records on its own SD card at the same interval.

The recorded variables include air temperature (°C), relative humidity (%), wind speed (km h$^{-1}$), wind direction (16-sector bearings), and rainfall (mm). This synchronized, high-resolution dataset provides the basis for thermal modeling and conservation planning by linking indoor microclimates with external weather conditions.

### 3.2 Empirical Corrosion Model

The corrosion rate was modelled using a physics-based equation derived from electrochemical principles, incorporating the Arrhenius relationship for temperature dependence and a power-law dependence on relative humidity. The formula is given by:

$$CR = C \times RH^n \times exp(\frac{-E_a}{R \times T}) \qquad (1)$$

where:

- *CR*: Corrosion Rate (μm/year)
- *C*: Calibrated coefficient
- *RH*: Relative humidity (as a fraction, e.g., 70% → 0.70)
- *n*: Humidity exponent (calibrated)
- *Ea*: Activation energy = 50,000 J/mol
- *R*: Universal gas constant = 8.314 J/(mol·K)
- *T*: Temperature in Kelvin = T (Celsius) + 273.15

### 3.2.1 Calibration Procedure

To ensure the model's accuracy, curve fitting procedure based on least-squares optimization was used to calibrate the model parameters *C* and *n* using literature data [5]. The calibration process involved the following steps:

- Input Data: Observed corrosion rates, relative humidity, and temperature from literature.
- Optimization Method: A least-squares optimization was performed to minimize the residuals between predicted and observed corrosion rates.
- Initial Guess: Initial values for C and n were set to 50.0 and 1.5, respectively, based on preliminary analysis.
- Residual Function: The residual was defined as the difference between predicted and observed corrosion rates:

$$Residual = C \times RH^n \times exp(\frac{-E_a}{R \times T}) - CR\_observed \quad\quad (2)$$

The optimization yielded calibrated values for *C* and *n*, which were subsequently used for corrosion rate calculations.

### 3.3 Data Collection and Preprocessing

The dataset comprises three years of sensor-derived environmental measurements (2019–2022) collected from multiple monitoring stations to study atmospheric corrosion of metallic materials. The raw data, includes timestamped records of temperature (°C), relative humidity (%), and station identifiers, capturing a robust sample of environmental conditions influencing corrosion. After rigorous preprocessing, the final dataset consists of 280,967 entries, providing a statistically robust foundation for reliable model training, inference, and performance evaluation in corrosion studies.

Preprocessing ensured data quality and compatibility with a physics-based corrosion model. Timestamps were standardized to a consistent format and converted to datetime objects, with invalid entries removed. Relative humidity values were clipped to a physically realistic range of 0–100% to eliminate outliers. Missing temperature and humidity values were imputed using forward and backward filling within each station to preserve temporal continuity. Temperature was converted to Kelvin, and relative humidity was normalized to a 0–1 scale for model input. A corrosion rate (µm/year) was computed using a calibrated model.

To capture temporal dynamics, 24-hour rolling window features were computed, including mean and standard deviation of relative humidity, mean temperature (Kelvin), and hours with relative humidity >0.8 ("hours wet"), a key corrosion driver. Rows with missing values post-processing were excluded, yielding the final dataset.

*Figure 6* illustrates seasonal trends, showing monthly mean corrosion rate (µm/year), relative humidity (%), and temperature (°C), with ±1 standard deviation shaded regions. Elevated corrosion rates at the beginning of the year and towards the end, highlight the dataset's ability to capture environmental influences on corrosion, supporting its suitability for advanced analyses in atmospheric corrosion research. Also noted is the correlation between relative humidity and corrosion rate.

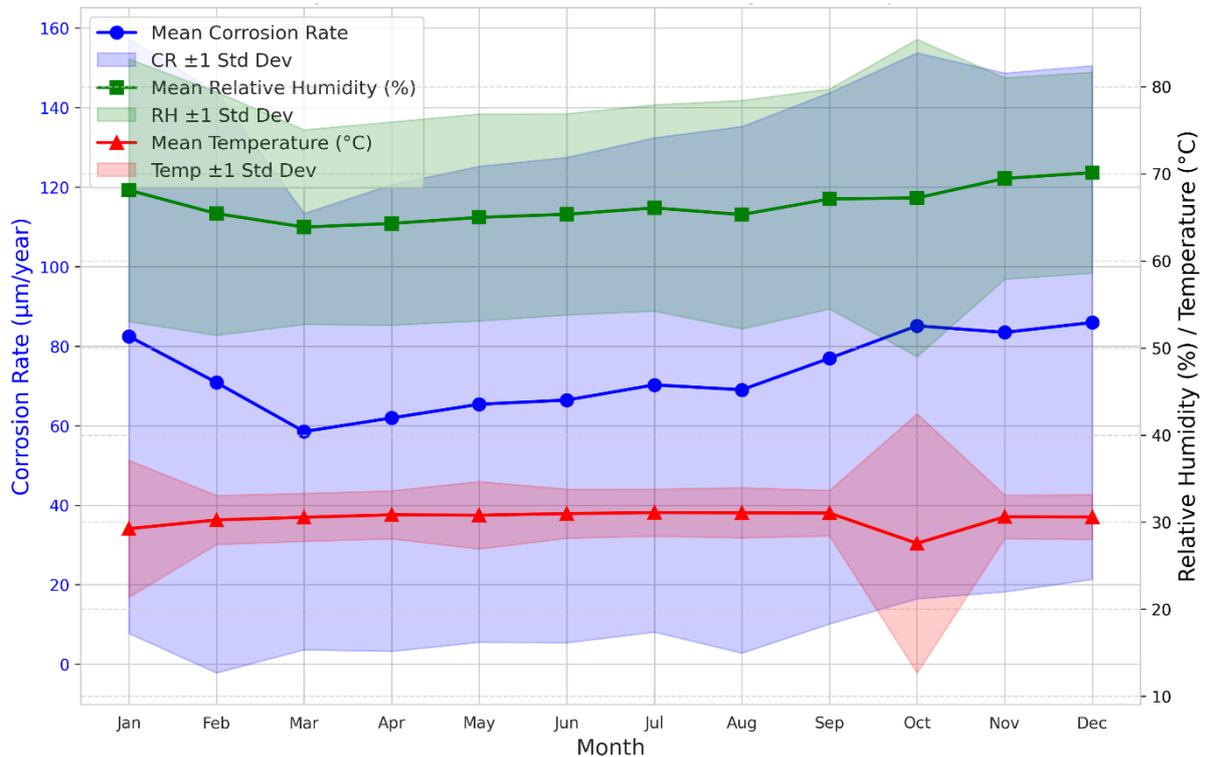

*Figure 3 Corrosion Rate, Relative Humidity and Temperature Across the year.*

### 3.4 Machine Learning Pipeline

This section delineates a machine learning pipeline for forecasting corrosion rates and categorizing corrosion risk levels based on environmental data, as extracted from the included preprocessing and training scripts. The pipeline includes feature engineering, model training for regression and classification tasks, and thorough assessment metrics, tailored to the stringent requirements of corrosion science research. The pipeline seeks to deliver precise forecasts and comprehensible insights for corrosion management in metallic structures by incorporating physics-informed features and sophisticated machine learning algorithms.

### 3.4.1 Feature Engineering

The feature engineering process transforms raw environmental data into a robust set of predictors tailored for corrosion rate prediction. The input dataset, generated from a preprocessing step, includes temperature and humidity measurements, critical drivers of corrosion due to their influence on reaction kinetics (via the Arrhenius equation) and surface moisture availability, respectively, which govern electrochemical corrosion mechanisms.

From the dataset of 280,967 samples, 9 numeric features were selected, excluding the target variable, timestamp, and temporal derivatives to prevent target leakage and reduce dimensionality. Missing values were imputed using feature-specific means to ensure data completeness. Features were standardized using StandardScaler to achieve a mean of zero and unit variance, facilitating compatibility with models such as Linear Regression, Random Forest Regressor, Gradient Boosting Regressor, and XGBoost Regressor. To capture temporal dynamics inherent in corrosion processes, the dataset was sorted by timestamp, enabling time-series-aware modeling.

### 3.4.2 Model Training

The regression models for predicting corrosion rates were trained using a time-series split to preserve the temporal sequence of the data, ensuring training data preceded test data to emulate real-world forecasting conditions. The dataset, comprising 280,967 samples with 9 features, was split into a 75/25 train-test ratio (224,773 training samples and 56,194 test samples) to maximize data utilization. Features were standardized using StandardScaler, fitted on the training set and applied separately to the test set to prevent data leakage. Linear Regression served as a baseline model, while Random Forest Regressor, Gradient Boosting Regressor, and XGBoost Regressor were utilized to capture non-linear corrosion patterns. Hyperparameter optimization was performed using Grid Search, tuning parameters such as learning rates (0.01–0.1) for boosting models and tree-based parameters (e.g., max_depth, n_estimators) for Random Forest, with internal cross-validation to select optimal configurations, ensuring robust model performance and computational efficiency.

### 3.4.3 Evaluation Metrics

To assess the performance of regression models in predicting corrosion rates (in μm/year), three complementary metrics were employed: Mean Squared Error (MSE), Root Mean Squared Error (RMSE), and Mean Absolute Error (MAE), alongside the Coefficient of Determination ($R^2$).

Mean Squared Error (MSE): MSE quantifies the average of the squared differences between predicted and actual corrosion rates, calculated as:

$$MSE = \frac{1}{n}\sum_{i=1}^{n}(y_i - \hat{y}_i)^2 \quad\quad\quad\quad (3)$$

where $y_i$ is the actual corrosion rate, $\hat{y}_i$ is the predicted rate, and $n$ is the number of samples. Expressed in (μm/year)², MSE emphasizes larger prediction errors due to squaring, providing a sensitive measure of model accuracy for corrosion rate forecasting, where precise predictions are critical for material degradation assessment.

Root Mean Squared Error (RMSE): RMSE, the square root of MSE, is calculated as:

$$RMSE = \sqrt{MSE} \quad\quad\quad\quad (4)$$

It measures the standard deviation of prediction errors in the original units of corrosion rate (μm/year). It offers an interpretable measure of typical prediction error magnitude, balancing sensitivity to outliers with practical relevance for evaluating model performance in corrosion studies.

Mean Absolute Error (MAE): MAE computes the average of absolute differences between predicted and actual corrosion rates, defined as:

$$MAE = \frac{1}{n}\sum_{i=1}^{n}|y_i - \hat{y}_i| \quad\quad\quad\quad (5)$$

Measured in μm/year, MAE provides a robust estimate of average prediction error, less sensitive to outliers than MSE or RMSE, making it valuable for assessing model reliability across diverse corrosion scenarios.

Coefficient of Determination ($R^2$)

R² evaluates the proportion of variance in the actual corrosion rates explained by the model, calculated as:

$$R2 = 1 - \frac{SSE}{SST} \quad\text{_____________(6)}$$

where SSE is Sum of Squares Error and SST is Total Sum of Squares. Ranging from 0 to 1, higher R² values indicate a better fit, with values above 0.8 typically reflecting strong predictive capability for regression models in corrosion rate prediction, capturing the complex interplay of environmental factors.

These metrics collectively provide a robust framework for evaluating model performance, balancing sensitivity to errors (MSE, RMSE), robustness to outliers (MAE), and explanatory power (R²), ensuring comprehensive assessment of predictive accuracy for corrosion rate forecasting.

## 4.0 Results and Discussion

### 4.1 Machine Learning Model Performance

A comparative assessment of the three regression models in *Figure 7 and Figure 8* highlighted distinct trade-offs between predictive accuracy and computational efficiency. Random Forest consistently outperformed the other models in terms of predictive accuracy, achieving the lowest overall prediction errors across all metrics. This indicates its strong capacity to capture complex nonlinear relationships in the corrosion data. However, this superior accuracy came at the cost of substantially longer training times, reflecting the computational intensity of constructing a large ensemble of decision trees. Despite this, its inference time remained practical, making it well-suited for applications where model training can be performed offline and prediction speed is prioritized.

Gradient Boosting demonstrated a balanced performance profile. While slightly less accurate than Random Forest, it maintained competitive predictive capability with moderate error values. Its training time was shorter than that of Random Forest, and it achieved the fastest inference speed among the three models. This balance of accuracy and efficiency suggests that Gradient Boosting may be advantageous in scenarios where rapid deployment and near-real-time predictions are critical.

XGBoost offered the most efficient computational profile, with exceptionally fast training and near-instantaneous inference. Although its predictive accuracy was marginally lower than that of the other two models, it remained within a high-performance range. This makes XGBoost particularly attractive for applications requiring rapid retraining or frequent model updates, where computational efficiency outweighs the slight trade-off in accuracy.

Overall, these findings underscore that Random Forest is the optimal choice when predictive precision is paramount, while XGBoost emerges as the most suitable model for scenarios prioritizing computational speed and scalability. Gradient Boosting, meanwhile, provides a versatile middle ground between the two, offering a pragmatic balance of accuracy and efficiency.

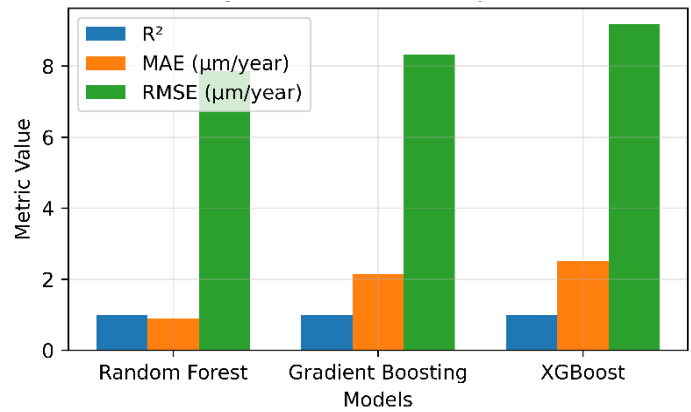

Figure 7 Model Performance Comparison

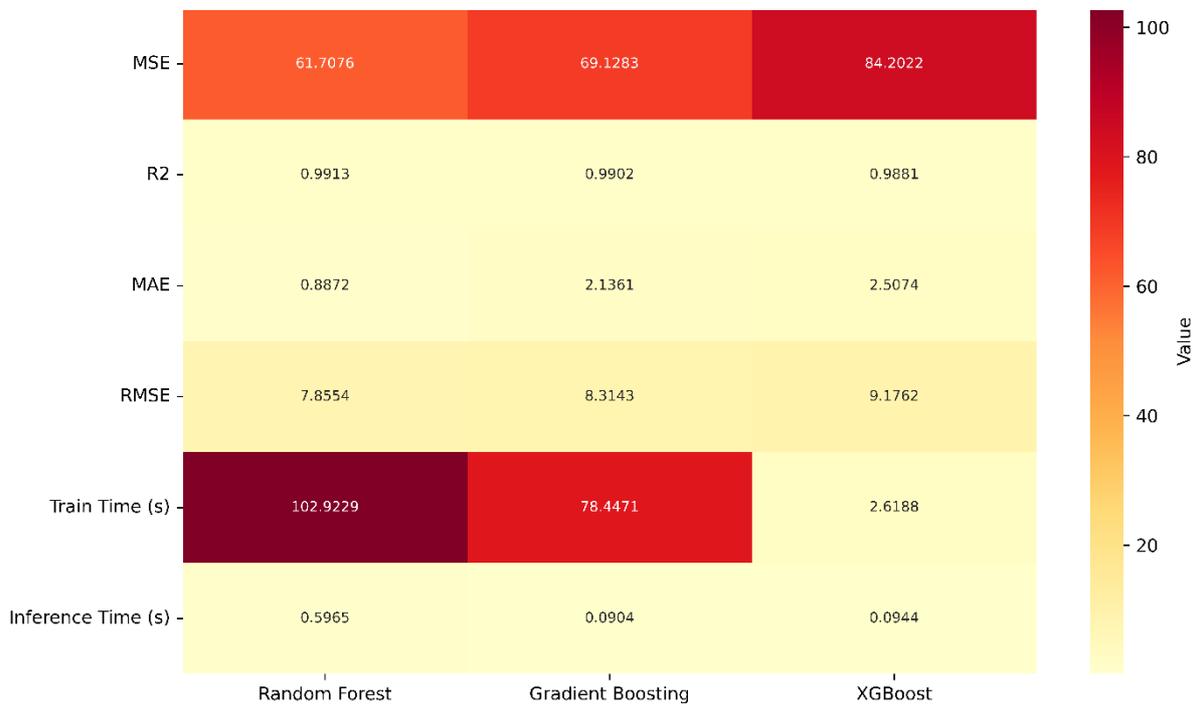

Figure 8 Regression Results

### 4.2 Model Deployment and Simulation

The Streamlit-based web application utilizes pre-trained machine learning models for real-time corrosion monitoring, forecasting corrosion rates and risk levels for metallic structures such as the San Sebastian Basilica. It utilizes trained models stored as .pkl files, along with a scaler and a feature list. Models and scalers are loaded using joblib, guaranteeing efficient loading. Users choose regression and classification models through sidebar dropdown menus, including a contingency simulation mode for reliability in the event that models are inaccessible. A public ngrok tunnel exposes the application on port 8501, facilitating remote access for practical utilization and providing real-time corrosion information with low latency.

The simulation framework, serving as a proof of concept, illustrates model capability without real-time sensor data, utilizing statistics from the acquired data (mean humidity 66.5% with std 13.2%; mean temperature: 30.3°C with std 5.7°C) to replicate genuine conditions. It operates in a continuous loop, refreshing every 2 seconds, producing time-series data for temperature, humidity, corrosion rate, and risk level. The corrosion rate function integrates a baseline rate with the influences of humidity and temperature, which are intensified during specific events. Mitigation measures, such as dehumidification, initiate a gradual healing process. Interactive recommendations and visualizations improve interpretability, hence proving the algorithms predicted efficacy for corrosion monitoring.

The charts clearly show the system's ability to forecast corrosion trends. *Figure 8* displays normal operation where both environmental conditions and corrosion rates stay within safe limits. *Figure 9* shows how the system detects early warning signs when humidity and/or temperature rise, predicting the coming increase in corrosion rates. Most importantly, *Figure 10* demonstrates how the system recognizes when corrosion risks are decreasing after implementing solutions like dehumidification, giving users confidence that their protective measures are working. This approach represents a significant improvement over traditional corrosion monitoring that typically only detects problems after they occur. By providing early warnings and confirming when solutions are effective, the system enables maintenance teams to take preventive action before damage occurs, potentially saving substantial repair costs and extending the life of heritage structures.

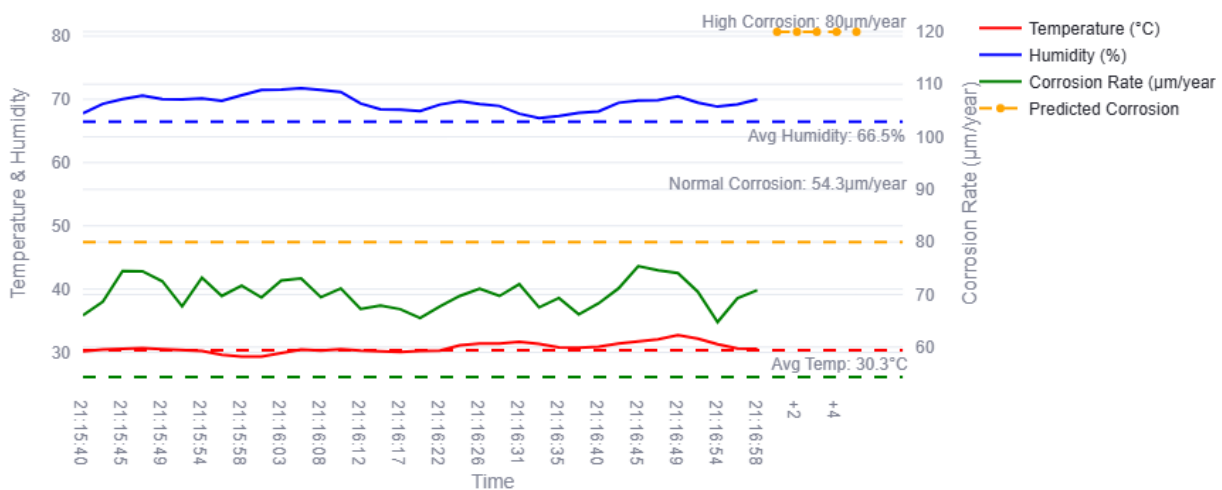

*Figure 8 Normal Measurement*

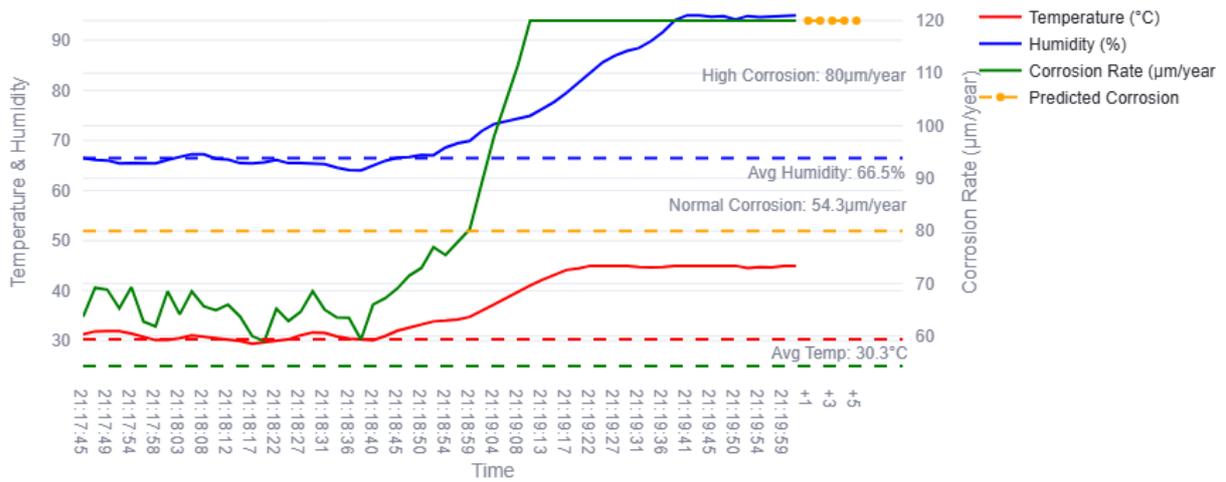

*Figure 9 Corrosion Predicted to Increase*

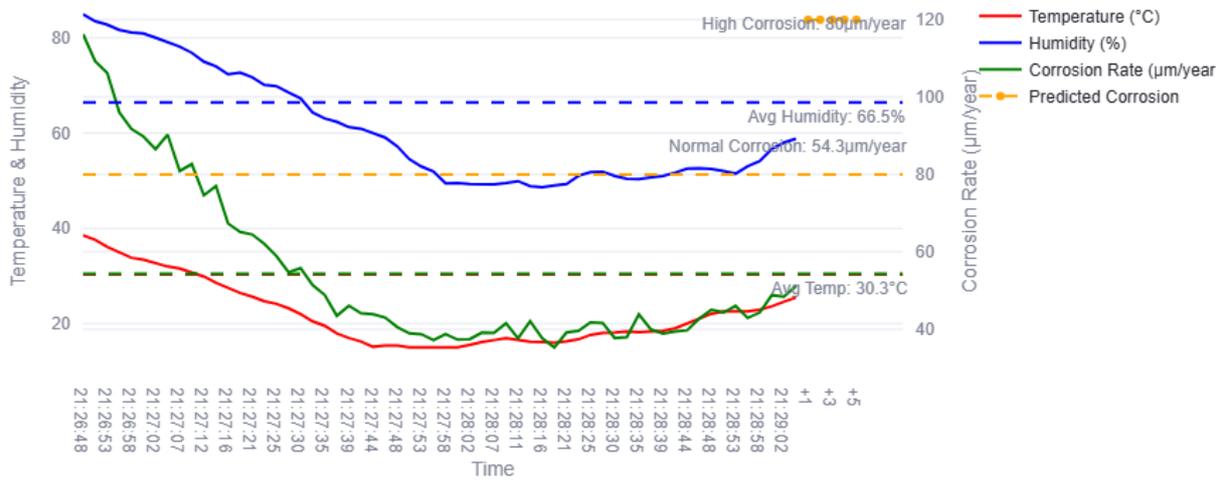

*Figure 4 Corrosion Predicted to Return to Normalcy*

### 4.3 Insights into Heritage Conservation

The regression outcomes provide essential instruments for heritage conservation, facilitating accurate corrosion rate forecasts and risk evaluations to safeguard cultural heritage edifices. The Random Forest Regressor surpasses both Gradient and XGBoost in delivering precise corrosion rate estimations. The exceptional precision of Random Forest regression facilitates the identification of high-risk zones in heritage structures, warranting prompt inspections and interventions. This corresponds with the assertions of [1], who endorse IoT-enabled monitoring for scalable heritage evaluations. The impeccable recall of the categorization models guarantees that no high-risk situations are overlooked, which is essential for safeguarding valuable assets, as highlighted by [9] in IoT–AI frameworks. In contrast to the DRFs proposed by [3], which depend on pollutant data frequently inaccessible in heritage contexts, our models utilize globally quantifiable temperature and humidity, hence improving applicability. The Streamlit dashboard's real-time alerts and recommendations (e.g., "activate dehumidifiers") facilitate preventive interventions, such as applying protective coatings during anticipated high-humidity occurrences, addressing resource limitations as identified in CULTCOAST [8].

The regression models' nearly flawless R² values provide dependable predictions of corrosion rates, guiding maintenance schedules and environmental regulations for heritage sites. This supports the findings of [13], who employed hybrid machine learning for limited corrosion data, and [14], who recognized temperature and humidity as primary factors. In contrast to the spatially constrained DRFs identified by [4] and [5], our models exhibit generalizability across various settings by emphasizing climate characteristics, making them suitable for heritage sites with inadequate pollution monitoring. The dashboard's time-to-failure visualizations facilitate lifespan estimations, aiding in strategic planning, while categorization warnings advise investments in temperature control, such as ventilation systems, in accordance with [7]. Nonetheless, flawless classification scores indicate possible overfitting, requiring validation with empirical data to confirm robustness, as emphasized by [10] regarding interpretable machine learning in maritime settings.

The pipeline exceeds conventional DRFs by utilizing temperature and humidity, tackling heritage-specific deficiencies in scalability and little intervention [9]. It facilitates focused inspections and enduring conservation efforts; nevertheless, additional validation is required to address class imbalance risks and improve practical applicability.

## 5.0 Conclusion

This study presents an integrated framework for predictive corrosion monitoring in cultural heritage conservation, combining innovative IoT data acquisition with physics-informed machine learning. We developed a custom LoRa-based wireless sensor network that enables minimally invasive, long-range monitoring of critical environmental parameters including temperature, humidity, wind speed, and direction without compromising structural integrity. We further established a physics-based corrosion model calibrated against empirical data, demonstrating that accurate degradation forecasting can be achieved using minimal environmental inputs. Finally, we introduced a dual-output predictive framework that integrates high-precision regression for long-term corrosion estimation with a binary classification system for immediate risk alerts, providing both strategic planning tools and actionable conservation guidance.

Our approach successfully bridges the gap between theoretical corrosion science and practical heritage preservation, offering a scalable, cost-effective solution that operates effectively even in resource-constrained environments. The system's ability to function with sparse sensor data while maintaining high predictive accuracy addresses a critical challenge in real-world cultural heritage monitoring, where complete and clean datasets are often unavailable.

Several promising directions emerge from this research. First, we intend to develop a robust data preprocessing pipeline specifically designed to handle missing and corrupted sensor data addressing a common practical challenge in long-term environmental monitoring. This pipeline will incorporate advanced imputation techniques and anomaly detection algorithms to enhance data quality before model ingestion. Second, we plan to expand validation across diverse heritage sites with varying materials, climates, and architectural configurations to assess the generalizability and robustness of our models under different environmental factors. Furthermore, we will explore the integration of additional sensor modalities, such as airborne pollutant concentrations and surface moisture measurements, to enhance model comprehensiveness while retaining cost-effectiveness.